\documentclass[12pt,letterpaper]{article}
\usepackage{amsfonts}
\usepackage{amsmath}
\usepackage{amssymb}
\usepackage{bm}
\usepackage{array}
\usepackage{geometry}
\usepackage{setspace}
\usepackage{graphicx}
\usepackage{lscape}
\usepackage{verbatim}
\usepackage{natbib}
\usepackage{color}
\usepackage{xcolor}
\usepackage{tabularx}
\definecolor{winered}{rgb}{0.5,0,0}
\usepackage[bookmarks=true, bookmarksnumbered=true, allbordercolors={1 1 1}]{hyperref}
\hypersetup{
  colorlinks   = true, 
  urlcolor     = blue, 
  linkcolor    = blue, 
  citecolor    = winered,
}
\usepackage[open=true, numbered=true]{bookmark}
\usepackage{float}
\usepackage{fancyhdr}
\usepackage[toc,page]{appendix}
\usepackage{scalefnt}
\usepackage{afterpage}
\usepackage[left]{lineno}
\usepackage{caption}
\usepackage{subcaption}
\usepackage{longtable}
\usepackage{authblk}
\usepackage{comment}

\makeatletter
\def\blfootnote{\xdef\@thefnmark{}\@footnotetext}
\makeatother

\emergencystretch=\maxdimen
\begin{document}
\title{\vspace{-1 cm} \huge{Exploring Monetary Policy Shocks with Large-Scale Bayesian VARs}\blfootnote{\hspace{-0.7cm} I would like to thank Martin Bruns, Luca Gambetti, Domenico Giannone, Michele Lenza, Nicol\`{o} Maffei-Faccioli, Mirela Miescu, Ivan Petrella, Giorgio Primiceri, Barbara Rossi and Lorenza Rossi for their valuable comments and suggestions. I also thank participants at the University of Lancaster's Workshop on Empirical and Theoretical Macroeconomics, the University of East Anglia ``2nd Time Series Workshop'', the Collegio Carlo Alberto conference on ``The Economics of Risk: Econometric Tools and Policy Implications'', and seminar participants at Universities of Manchester and Paris Dauphine for their insightful feedback. Any remaining errors are solely my responsibility.
\\
\indent Correspondence: Professor of Econometrics, Adam Smith Business School, University of Glasgow, 2 Discovery Place, Glasgow, G11 6EY, United Kingdom; email: \href{mailto:dikorobilis@googlemail.com}{dikorobilis@googlemail.com}.
}}
\author[1,2,3]{Dimitris Korobilis}
\affil[1]{\vspace{-0.1 cm}{\footnotesize \emph{University of Glasgow, UK}}}
\affil[2]{\vspace{-0.1 cm}{\footnotesize \emph{Center for Applied Macroeconomics and Commodity Prices (CAMP), Norway}}}
\affil[3]{\vspace{-0.1 cm}{\footnotesize \emph{Rimini Center for Economic Analysis (RCEA), Italy}}}
\date{\today}

\maketitle

\begin{abstract}
\noindent I introduce a high-dimensional Bayesian vector autoregressive (BVAR) framework designed to estimate the effects of conventional monetary policy shocks. The model captures structural shocks as latent factors, enabling computationally efficient estimation in high-dimensional settings through a straightforward Gibbs sampler. By incorporating time variation in the effects of monetary policy while maintaining tractability, the methodology offers a flexible and scalable approach to empirical macroeconomic analysis using BVARs, well-suited to handle data irregularities observed in recent times. Applied to the U.S. economy, I identify monetary shocks using a combination of high-frequency surprises and sign restrictions, yielding results that are robust across a wide range of specification choices. The findings indicate that the Federal Reserve's influence on disaggregated consumer prices fluctuated significantly during the 2022–24 high-inflation period, shedding new light on the evolving dynamics of monetary policy transmission.

\bigskip


\bigskip \medskip

\end{abstract}
\thispagestyle{empty} 

\newpage
\doublespacing

\setcounter{page}{1}
\section{Introduction}

This paper quantifies the effectiveness of conventional monetary policy shocks during the post-pandemic inflationary period using large-scale vector autoregressions (VARs). While previous studies \citep[see, for example,][]{Banburaetal2010,BernankeBoivin2003,Giannoneetal2015} have established the benefits of identifying monetary policy in data-rich environments, this research develops a flexible yet numerically robust VAR framework to address emerging estimation and identification challenges. First, after a prolonged period of near-zero interest rates, historical data may be less informative for evaluating recent monetary policy effectiveness.\footnote{Studies examining conventional monetary policy (MP) shocks, such as \cite{Ariasetal2019,Read2024,Schlaak2023}, rely on data ending in 2007, that is, just before the period when nominal interest rates effectively reached their zero lower bound. \cite{AruobaDrechsel2024} perform their empirical analysis to 2008, arguing that their findings remain relevant for the 2022–2023 inflationary period, provided that the transmission mechanism of MP shocks has not changed. However, they caution that factors such as state dependency, non-linearities, and structural shifts may alter response dynamics and invalidate projections to the recent period.} Second, the COVID-19 pandemic introduced substantial outliers, disrupting macroeconomic correlations and complicating structural identification.\footnote{\cite{LenzaPrimiceri2022} and \cite{CarrieroClarkMarcellinoMertens2024} develop VAR methods to accommodate COVID‑19 outliers and stochastic volatility; they do not, however, address structural shock identification or inference in high-dimensional systems.} Third, in an increasingly globalized and financially volatile environment, enhancing the VAR with relevant predictors is crucial for statistical fit and for ensuring that structural shocks are well identified. Moving forward, addressing these data challenges requires novel statistical methods and estimation algorithms, capable of handling high-dimensional data, nonlinearities, and other complexities. A major contribution of this study is the development of such methods, allowing the integration of high-dimensional predictors to the application of robust identification techniques for structural shocks.

To tackle these challenges, I develop a high-dimensional and data-analytic approach that integrates economic identification schemes with a Bayesian shrinkage estimation methodology. Building on \cite{Korobilis2022}, I extend the VAR to integrate high frequency instruments, fat-tailed errors, time-varying parameters, and stochastic volatility. The approach uses ``target'' and ``path'' factors of high frequency surprise series from \cite{Gurkaynaketal2005} to look at how the markets react to Federal Reserve Bank's policy announcements. The target factor describes the effects of surprise changes in the current federal funds rate target, that is, conventional shocks that are of interest in this paper. The path factor captures the future policy path which is used to identify a residual monetary policy shock.\footnote{There are varying views about the number and effects of future policy path and unconventional shocks, in general. \cite{NakamuraSteinsson2018} argue that the Fed has private information (what they call the ``Fed information effect''), while \cite{BauerSwanson2023} argue that both the Fed and professional forecasters respond to incoming economic news, which naturally generates a positive correlation between monetary policy surprises and forecast revisions, without requiring the Fed to have private information about the economy. \cite{Swanson2021} extends \cite{Gurkaynaketal2005} to include a third, quantitative easing factor, and \cite{Jarocinski2024} identifies up to three unconventional monetary policy shocks.} I first show that in a small-scale VAR the use of high-frequency surprises as internal instruments for identifying monetary policy is not always robust and in line with theory \citep[see also][for similar arguments]{Acosta2023,Lakdawala2019}.\footnote{Several other studies have raised concerns about the use of high-frequency instruments (HFIs) in identifying monetary policy shocks, particularly regarding issues of exogeneity, temporal aggregation bias, and event window limitations; see \cite{BauerSwanson2023}, \cite{paccagnini2021identifying} and \cite{casini2024limitations}, among others.} 

Next, I show that a plausible solution to this lack of identification is a hybrid method that merges proxies with zero and sign restrictions to the response of key macroeconomic aggregates -- namely measures of output, inflation, short/medium/long interest rates, and money supply. While sign restrictions to these variables align with conventional theoretical predictions of a monetary policy shock, they alone are also insufficient for unique identification and thus need to be combined with high-frequency instruments.\footnote{\cite{BraunBrueggemann2023} present similar arguments in the context of VARs that combine external instruments with sign restrictions. While their algorithm is very useful, it is not computationally feasible for large VARs, particularly when allowing for time-varying shock impacts.} An emerging line of research, such as \cite{Acosta2023}, \cite{BraunBrueggemann2023}, \cite{Carrieroetal2024}, \cite{Read2024} and \cite{Schlaak2023}, demonstrates the advantages of combining established identification schemes, such as sign, zero and narrative restrictions, proxies, and identification via heteroskedasticity and non-Gaussianity. \cite{Prueser2024} and \cite{chan2022large} similarly extend the VAR methodology of \cite{Korobilis2022} by adapting it to identification by non-Gaussianity and heteroskedasticity, respectively. However, these identification approaches are statistical in nature, and may not always ensure adequate identification of conventional monetary policy shocks. In contrast, I provide comprehensive evidence that combining high-frequency instruments with sign restrictions remains economically robust across various data and modeling assumptions, including lag length, stationarity transformations of variables, assumptions about residual shocks, the measurement of short-term interest rates, and different assumptions regarding the exogeneity of instruments.

After establishing that the hybrid identification scheme performs effectively in a small-scale VAR, the analysis examines the impact of post-pandemic conventional monetary policy on the broader economy, with a particular focus on disaggregated components of consumer inflation. The VAR model is expanded to include key financial indicators, uncertainty measures, expectations, and other predictors, ensuring that the residuals effectively capture structural monetary policy shocks. These predictors and inflation components are permitted to exhibit time-varying responses to monetary policy shocks, accommodating changes in the transmission mechanism over time, with idiosyncratic shocks modeled to incorporate stochastic volatility dynamics. Employing Bayesian shrinkage and variable selection techniques allows for the semi-automatic determination of the optimal degree of time variation within the system, adeptly adjusting to abrupt changes, such as those introduced by post-COVID-19 outliers. This approach differs from \cite{MumtazPetrova2023}, where all parameters in their proxy VAR are time-varying. In contrast, our methodology maintains constant weights for high-frequency instruments and key macroeconomic aggregates (those subjected to sign restrictions) when loading onto structural shocks. This strategy ensures robust identification through these variables while permitting other variables of interest to have unrestricted, time-varying responses to monetary policy shocks. Consequently, the large-scale system integrates multiple identification layers, balancing the imposition of economic theory with the flexibility for data to adapt to abrupt regime shifts experienced since the global financial crisis.

The proposed VAR methods are applied to monthly macroeconomic data for the US for the period 1995-2024. Among the large number of variables, the variables of interest are disaggregate price indices and their response to conventional monetary policy shock during the post-pandemic inflationary period. Based on the definitions of the Bureau for Labor Statistics, I use eight major categories of the US consumer price index (CPI) such as food, apparel, and education, in order to pin down the heterogeneous effect of policy shocks. I find that the effects of monetary tightening vary substantially across CPI components, with core goods exhibiting rapid and pronounced disinflationary responses, while services (particularly housing) adjust more sluggishly and with long lags. Importantly, these sector-specific responses have evolved over time: in the post-pandemic period, the impact of conventional policy shocks on goods prices has intensified, likely reflecting the unwinding of pandemic-era demand and supply imbalances. Despite this heterogeneity, virtually all components eventually move in the direction of lower inflation following a contractionary shock, confirming that monetary policy remains a broadly effective tool for influencing inflation dynamics, even in a high-inflation environment.

The paper is organized as follows. The next Section presents the Bayesian VAR methodology and builds the various components that will be relevant for large-scale estimation and structural inference. Section 3 details the variables, sources and transformations used in this high-dimensional setting. In Section 4 I provide the empirical results, which are also complemented by an online supplement. Section 5 concludes the paper.

\section{Econometric methodology}
The starting point is the $p$-lag vector autoregression of the form
\begin{equation}
\bm y_{t} = \bm \phi_{0} + \sum_{j=1}^{p} \bm \Phi_{j} \mathbf{y}_{t-j} + \bm \varepsilon_{t}. \label{VARp}
\end{equation}
This can be written more compactly as
\begin{equation}
\bm y_{t} = \bm \Phi \bm x_{t} + \bm \varepsilon_{t}, \label{VAR}
\end{equation}
where $\bm y_{t}$ is an $\left( n \times 1 \right)$ vector of observed macro variables, $\bm x_{t} = \left( 1,\bm y_{t-1}^{\prime},...,\bm y_{t-p}^{\prime} \right)^{\prime}$ a $\left( k \times 1 \right)$ vector of lags and the intercept, $k=np+1$, $\bm{\Phi} = \left[\bm{\phi}_{0},\bm \Phi_{0},...,\bm \Phi_{p} \right]$ is an $(n \times k)$ matrix of coefficients, and $\bm{\varepsilon}_{t}$ a $\left( n \times 1 \right)$ vector of disturbances distributed as $N\left( \bm{0}_{n \times 1},\bm{\Omega} \right)$ with $\bm{\Omega}$ an $n \times n$ symmetric and positive semi-definite covariance matrix. The structural VAR (SVAR) form is derived by left-multiplying \eqref{VARp} with $\bm A$ yielding
\begin{equation}
\bm A \bm y_{t} = \bm B \bm x_{t} + \bm u_{t}, \label{SVAR}
\end{equation}
where $\bm A^{-1} \bm A^{-1\prime} = \bm \Omega$ is a decomposition of the reduced-form covariance matrix, $\mathbf{B} = \bm A \mathbf{\Phi}$ are the SVAR coefficients and $\bm u_{t} \sim N(\bm 0,\bm I)$ are structural shocks where it holds that $\bm \varepsilon_{t} = \bm A^{-1}  \bm u_{t}$.

The VAR in \eqref{VARp} can be estimated with standard tools, providing estimates of $\widehat{\bm \Phi}$ and $\widehat{\bm \Omega}$. However, recovering the matrix $\bm A$ of structural contemporaneous restrictions is not feasible without further assumptions. $\bm A$ has $n^2$ elements, while $\widehat{\bm \Omega}$ has $n(n+1)/2$ elements estimated from data. Identification restrictions, usually stemming from economic theory, stylized facts, or intuition, fill the gap in providing information on the remaining $n(n-1)/2$ coefficients. I focus on two popular approaches to identification, namely sign restrictions and high frequency instruments. The former helps identify elements of $\bm A$ that satisfy expected signs; for example, if the response of variable $i$ to a shock in $j$ is positive, $i,j=1,...,n$, then $A_{ij}>0$. The latter approach uses $m$ proxies or instruments $m_{t}$ that are correlated with $m$ shocks of interest and uncorrelated with other shocks. In this paper, $m_{t}$ are high-frequency surprises in interest rate markets, so the terms proxies, instruments, and surprises may be used interchangeably.

\subsection{A large monetary policy Bayesian VAR}
In the presence of $m$ high frequency surprises $\bm m_{t}$ and large-$n$ macro variables $\bm y_{t}$, the VAR of \eqref{VAR} has to be augmented with the surprise data. Using the result above that $\bm \varepsilon_{t} = \bm A^{-1}  \bm u_{t}$, I define the following generic representation of this augmented VAR \citep[see also][]{BraunBrueggemann2023}
\begin{equation}
\left[
\begin{array}{c}
\bm  m_{t} \\
\bm  y_{t}
\end{array}
\right] =
\left[ 
\begin{array}{cc}
\bm \Phi^{mm} & \bm \Phi^{my}  \\
\bm \Phi^{ym} & \bm \Phi^{yy} 
\end{array}
\right]
\left[
\begin{array}{c}
\bm  m_{t-1} \\
\bm  y_{t-1}
\end{array}
\right]
+ \left[
\begin{array}{c}
\bm \Gamma \\
\bm A^{-1} 
\end{array}
\right]
\bm u_{t} + 
 \left[
\begin{array}{c}
\bm W^{1/2}  \\
\bm 0
\end{array}
\right]
\bm \eta_{t}
, \label{MPVAR}
\end{equation}
where $\bm \Phi^{ij}$ are the autoregressive coefficients of lags of variables $j$ in equations of variables $i$, with $i,j=\bm m, \bm y$, $\bm u_{t}  \sim N(\bm 0,\bm I)$ still denotes the structural VAR shock, the term $\bm \eta_{t}  \sim N(\bm 0,\bm I)$ is an idiosyncratic disturbance with $\bm \eta_{t} \bot \bm u_{t}$, $\bm \Gamma$ is a matrix of coefficients and $\bm W$ a diagonal matrix of variances. $\bm \Gamma$ measures the strength of the instrument's correlation with the structural shock, while $\bm W$ accounts for the possibility that the instrument is imperfect, allowing for measurement error or partial relevance. For the sake of notational simplicity the intercept term is excluded in the specification above, and I focus on the one-lag specification without loss of generality. \cite{MertensRavn2013} and \cite{JarocinskiKaradi2020} assume that $\bm \Phi^{mm} = \bm \Phi^{my} =\bm 0$. \cite{BauerSwanson2023} argue that interest rate surprises might be driven by macroeconomic news which is equivalent to assuming $\bm \Phi^{my} \neq 0$, as long as $\bm y$ also includes measures of macro news. Another empirically relevant question is also whether macroeconomic variables are dynamically independent of high frequency surprises, implying $\bm \Phi^{ym} = \bm 0$.

In higher dimensions, when $n$ is sufficiently large, it is expected that the VAR system does not necessarily have $n$ structural or primitive shocks. Instead, a smaller number of common, economically meaningful, driving forces are responsible for causing disruptions in the system. For that reason I follow previous work in \cite{Korobilis2022} and replace the identity $\bm \varepsilon_{t} = \bm A^{-1}  \bm u_{t}$ implied by equation \eqref{SVAR} with the factor-like decomposition
\begin{equation}
\bm{\varepsilon}_{t} = \bm \Lambda \bm f_{t} + \bm v_{t},
\end{equation}
where $ \bm f_{t} \sim N( \bm 0, \bm I) $ is an $r \times 1$ vector of factors with $r \ll n$, $\bm \Lambda$ an $n \times r$ matrix of loadings, and $\bm v_{t} \sim N(\bm 0, \bm \Sigma)$ is a vector of idiosyncratic disturbances where $\bm \Sigma$ is a diagonal matrix of variances.
Under this factor decomposition of the reduced-form disturbances, the VAR augmented with instruments of equation \eqref{MPVAR} can now be written as
\begin{equation}
\left[
\begin{array}{c}
\bm  m_{t} \\
\bm  y_{t}
\end{array}
\right] = \left[ 
\begin{array}{cc}
\bm \Phi^{mm} & \bm \Phi^{my}  \\
\bm \Phi^{ym} & \bm \Phi^{yy} 
\end{array}
\right]
\left[
\begin{array}{c}
\bm  m_{t-1} \\
\bm  y_{t-1}
\end{array}
\right]
+ \left[
\begin{array}{c}
\bm \Gamma \\
\bm \Lambda 
\end{array}
\right]
\bm f_{t} + 
 \left[
\begin{array}{c}
\bm W^{1/2}  \\
\bm \Sigma^{1/2}
\end{array}
\right]
\bm \eta_{t}
, \label{largeMPVAR}
\end{equation}
where in the notation above I have used the fact that $v_{t} \equiv \bm \Sigma^{1/2} \bm \eta_{t}$.

What is the economic interpretation of the model above? A smaller number of $r$ structural shocks $\bm f_{t}$ exist in the VAR. In order to see this, define $\bm y_{t}^{\star} = \left[\bm m_{t}^{\prime},\bm y_{t}^{\prime} \right]^{\prime}$ and $\bm \Phi = \left[ 
\begin{array}{cc}
\bm \Phi^{mm} & \bm \Phi^{my}  \\
\bm \Phi^{ym} & \bm \Phi^{yy} 
\end{array}
\right]$, $\bm \Gamma^{\star} = \left[
\begin{array}{c}
\bm \Gamma \\
\bm \Lambda 
\end{array}
\right]$ and $\bm W^{\star} =  \left[
\begin{array}{c}
\bm W  \\
\bm \Sigma
\end{array}
\right]$, and solve for the SVAR form as follows
\begin{eqnarray}
\bm y_{t}^{\star} & = & \bm \Phi \bm y_{t-1}^{\star} + \bm \Gamma^{\star} \bm f_{t} + \bm W^{\star 1/2} \bm \eta_{t}, \Rightarrow  \label{VAR_full} \\
\bm A^{\star} \bm y_{t}^{\star} & = &  \bm B \bm y_{t-1}^{\star} +  \bm f_{t} + \bm A^{\star}  \bm W^{\star 1/2} \bm \eta_{t}, \label{SVAR_full} \\
\bm A^{\star} \bm y_{t}^{\star} & \approx &  \bm B \bm y_{t-1}^{\star} +  \bm f_{t}, \label{SVAR_approx}
\end{eqnarray}
where $\bm A^{\star} = \left( \bm \Gamma^{\star \prime} \bm \Gamma^{\star} \right)^{-1} \bm \Gamma^{\star}$ is an $r \times (n+m)$ matrix, $\bm B^{\star} = \bm A^{\star} \mathbf{\Phi}$. Compared to the standard SVAR form in equation \eqref{SVAR} with contemporaneous relationship matrix $\bm A$, equation \eqref{SVAR_approx} is a reduced-rank SVAR with $r$ structural relationships among the $n$ macro variables (plus the $m$ instruments). This SVAR formulation results from the fact that $\bm A^{\star} \bm W^{\star 1/2} \bm \eta_{t} \rightarrow 0$ as $n \rightarrow \infty$ \citep{Korobilis2022}. Therefore, assuming the effect of the disturbance term $\bm \eta_{t}$ vanishes asymptotically in the SVAR representation, $\bm f_{t}$ is approximately equivalent to the vector of structural disturbances $\bm u_{t}$. However, in the reduced-form formulation the effect of $\bm \eta_{t}$ remains important and it has to be estimated alongside other parameter. This term has a natural interpretation too. From a statistical perspective, idiosyncratic disturbances $\bm \eta_{t}$ are considered to arise due to measurement error, outliers and other irregularities in each individual time series. From an economic point of view, the idiosyncratic disturbance term corresponds to expectations errors, heterogeneous information sets, myopia and other forms of irrational behavior \citep{Gorodnichenko2005}.

The result above suggest that impulse response functions (IRFs) are able to reflect the propagation of economically interpretable structural shocks, even in the presence of heterogeneous and noisy time series. IRFs trace the dynamic effect of a structural shock $f_{jt}$ on the vector of observables $\bm y_{t}^{\star} = \left[ \bm m_{t}^{\prime}, \bm y_{t}^{\prime} \right]^{\prime}$ over time. These effects can be obtained by considering the vector moving average (VMA) representation of the system in \eqref{largeMPVAR}, which takes the form:
\begin{equation}
\bm y_{t}^{\star} = \bm \mu + \sum_{h=0}^{\infty} \bm \Psi_{h} \left( \bm \Gamma^{\star} \bm f_{t-h} + \bm W^{\star 1/2} \bm \eta_{t-h} \right), \label{VMA_factor}
\end{equation}
where $\bm \mu$ is a constant mean vector, $\bm \Psi_{h}$ are the VMA coefficients implied by the VAR companion form, and $\bm \Gamma^{\star}$ stacks the proxy coefficients $\bm \Gamma$ and the macroeconomic loadings $\bm \Lambda$. 

The impulse response at horizon $h$ to the $j$-th structural factor $f_{jt}$ is:
\begin{equation}
\text{IRF}_{h}^{(j)} = \frac{\partial \bm y_{t+h}^{\star}}{\partial f_{jt}} = \bm \Psi_{h} \bm \Gamma^{\star}_{\cdot j}, \label{IRF_j}
\end{equation}
where $\bm \Gamma^{\star}_{\cdot j}$ is the $j$-th column of $\bm \Gamma^{\star}$. In particular, the impact IRF (i.e., the contemporaneous effect) is given by:
\begin{equation}
\text{IRF}_{0}^{(j)} = \bm \Psi_{0} \bm \Gamma^{\star}_{\cdot j} = \bm \Gamma^{\star}_{\cdot j}. \label{impact_IRF}
\end{equation}
This result is immediate from the identity $\bm \Psi_0 = \bm I$ and demonstrates that the loading vector $\bm \Gamma^{\star}_{\cdot j}$ defines the contemporaneous response of each observable variable to structural factor $f_{jt}$. Therefore, any parametric restrictions imposed on the sign or magnitude of elements of $\bm \Gamma$ and $\bm \Lambda$ (i.e., $\bm \Gamma^{\star}$) directly translate into sign and shape restrictions on the impact responses of the observed series. This insight underpins the identification strategy employed in this paper. Specifically, the combination of $\bm{\Gamma}$ and $\bm{\Lambda}$ determines the direction and magnitude of the contemporaneous responses to the structural factors $\bm f_t$. In the context of an instrument-augmented VAR, the matrix $\bm{\Gamma}$ governs the impact of the factors on the instrument block $\bm m_t$, while $\bm{\Lambda}$ governs the impact on the macroeconomic variables $\bm y_t$. 

In this paper, I impose that $\bm{\Gamma}$ is a diagonal matrix with positive entries, ensuring that each instrument loads positively on its corresponding factor while maintaining orthogonality across instruments. This choice is natural given the economic interpretation of the instruments and facilitates factor identification without requiring an arbitrary normalization such as $\bm{\Gamma} = \bm{I}_r$. While this restriction uniquely identifies a rotation of the factors, it does not guarantee that the resulting impulse responses of the macroeconomic variables are economically meaningful. Therefore, I impose additional economically motivated sign restrictions on selected rows of $\bm{\Lambda}$, corresponding to the expected effects of monetary policy shocks on key aggregates such as output and inflation. This two-layer identification strategy ensures that the factors are both statistically identified through the instrument block and economically interpretable through their impact on macroeconomic variables. These sign restrictions are trivial to impose. \cite{Korobilis2022} shows that such restrictions translate into truncated normal posterior distributions that are trivial to sample from: identification is embedded directly into the conditional posteriors, ensuring that every draw automatically satisfies the imposed restrictions. This one-step identification procedure provides major computational advantages, allowing the method to scale efficiently to large VAR systems. An outline of the posterior sampler is given in \autoref{sec:Bayesian} and full technical details of the sampler are provided in the online supplement.

\subsection{Adding time-variation, stochastic volatility and fat-tailed errors}
Time-varying parameter models have gained lots of prominence in macroeconomics, especially since the global financial crisis. They allow to accommodate structural breaks in the data in a flexible way. Compared to Markov switching and other regime-switching methods, time-varying parameters are able to unveil the full transmission mechanism of shocks. For that reason, I propose to extend the VAR of the previous Section into the following time-varying specification
\begin{equation}
\left[
\begin{array}{c}
\bm  m_{t} \\
\bm  y_{t}
\end{array}
\right] = \left[ 
\begin{array}{cc}
\bm \Phi^{mm} & \bm \Phi^{my}  \\
\bm \Phi^{ym} & \bm \Phi^{yy} 
\end{array}
\right]
\left[
\begin{array}{c}
\bm  m_{t-1} \\
\bm  y_{t-1}
\end{array}
\right]
+ \left[
\begin{array}{c}
\bm \Gamma \\
\bm \Lambda_{t} 
\end{array}
\right]
\bm f_{t} + 
 \left[
\begin{array}{c}
\bm W^{1/2}  \\
\bm \Sigma_{t}^{1/2}
\end{array}
\right]
\bm \eta_{t}. \label{largeMPVAR2}
\end{equation}
where the vectorized forms of the matrix $\bm \Lambda_{t}$ as well as each diagonal element $\log \left( \Sigma_{ii,t}\right)$, $i=1,...,n$ follow random walk specifications of the form
\begin{eqnarray}
vec \left( \bm \Lambda_{t} \right) & = & vec \left( \bm \Lambda_{t-1} \right) + \bm \xi_{t},  \label{stateLam} \\
\log \left( \Sigma_{ii,t}\right) & = & \log \left( \Sigma_{ii,t-1}\right) + \delta_{i,t}, \text{ \ \ \ } \forall i=1,...,n,
\end{eqnarray}
where $\bm \xi_{t} \sim N \left( \bm 0, \bm Q \right)$ with $\bm Q$ an $n \times n$ covariance matrix, and $\delta_{i,t} \sim N(0,\omega_{i}^{2})$ with $\omega_{i}^{2}$ a scalar variance parameter, for $i=1,...,n$. 

This form of time-variation has been used consistently since the early macroeconomic work of the 1970s and it implies the prior belief that each time varying parameter at time $t$ is centered at its time $t-1$ value. It is important to emphasize that the extent to which the responses of macroeconomic variables vary over time is not determined arbitrarily, but is governed by the prior placed on the innovation variance matrix $\bm Q$. In particular, I adopt a hierarchical horseshoe prior for each diagonal element of $\bm Q$, which ensures adaptive shrinkage \citep{Carvalhoetal2010}. This prior is both automatic and tuning-free, and requires no manual calibration of shrinkage parameters. Each element of $\bm Q$ is decomposed into global and local variance components, which jointly control the degree of time variation in the corresponding loading coefficient. The prior shrinks small innovations in $\bm \Lambda_t$ aggressively toward zero, effectively removing time variation where not supported by the data. At the same time, it allows large innovations to persist, thereby capturing structural breaks. This regularization mechanism ensures that the impulse responses of disaggregated prices and other unrestricted variables are shaped primarily by the data, rather than by arbitrary assumptions about the nature or smoothness of time variation.

Time-varying parameters and stochastic volatility are important for capturing structural breaks. However, they may not effectively handle the large outliers associated with the COVID-19 period. For this reason, I decompose the idiosyncratic error $\bm \eta_{t}$ into two components: one corresponding to the proxies, $\bm \eta_{t}^{m}$, and another corresponding to the macroeconomic variables, $\bm \eta_{t}^{y}$. I assume:
\begin{equation}
\bm \eta_{t} = \left[ \begin{array}{c}
\bm \eta_{t}^{m} \\
\bm \eta_{t}^{y}
\end{array} \right] \sim \left( \begin{array}{c}
N(\bm 0_{m \times 1}, \bm I_m) \\
t(\bm \nu, \bm 0_{n \times 1}, \bm I_n)
\end{array}
\right)
\end{equation}
where $t(\bm \nu, \bm 0, \bm I)$ denotes the standardized Student-t distribution with $\bm \nu$ an $n \times 1$ vector indicating the degrees of freedom for each macro variable. In this specification, $\bm \nu$ is a parameter estimated from the data. Smaller values of $\bm \nu$ allow for fatter tails and lower kurtosis, thus accommodating extreme outliers in the data.

Two key aspects of the specification above should be emphasized. First, the autoregressive parameters are not time-varying. Given the interest in high-dimensional VARs, the matrix $\bm \Phi = \left[ 
\begin{array}{cc}
\bm \Phi^{mm} & \bm \Phi^{my} \\
\bm \Phi^{ym} & \bm \Phi^{yy}
\end{array}
\right]$ may contain tens of thousands of elements. Consequently, converting such a large number of parameters into time-varying values would lead to significant estimation error and bias.\footnote{The limited degrees of freedom in the data and the infeasibility of running the MCMC chain for a sufficient number of iterations make accurate estimation difficult in very high-dimensional settings.} Moreover, ensuring that each draw of the autoregressive coefficients corresponds to a stationary VAR would be computationally impractical. Empirical evidence from the macroeconomics literature provides an additional reason to avoid time variation in $\bm \Phi$. For example, \cite{SimsZha2006} find minimal evidence of variation in $\bm \Phi$ over time, despite finding clear evidence of time-varying volatilities in a VAR. Even when $\bm{\Phi}$ is time-invariant, it can contain thousands of elements in high-dimensional settings. To regularize the estimation of this matrix, I follow \cite{Korobilis2022} and impose a horseshoe prior.

The second important aspect of the proposed time-varying extension is that the proxy equation parameters $\bm \Gamma$ and $\bm W$ should also remain constant, and the idiosyncratic term $\bm \eta_{t}^{m}$ should remain Gaussian. In the context of an external instruments VAR model, \cite{MumtazPetrova2023} allow these parameters to be time-varying. However, in the proposed specification, the internal instruments load on $\bm f_{t}$, a vector of latent time-varying parameters that must be estimated from the data. If $\bm \Gamma$ and $\bm W$ were also time-varying, they would absorb much of the informative content of the proxies, making the latent factors ``flat'' and unidentifiable. The same issue would arise if $\bm \eta_{t}^{m}$ followed a Student-t distribution: it would absorb many of the spikes in the surprise proxies, rendering the informational content of the factors economically irrelevant. In this model, a lack of factor identification translates into a failure to identify structural shocks, including the monetary policy shock of interest. Therefore, it is more appropriate to treat the relevance of the proxies as constant and the idiosyncratic shocks as Gaussian, while allowing the effects of identified shocks on macroeconomic variables to vary over time. Maintaining constant relevance of the proxies ensures that the model can more accurately and robustly identify the time-varying effects of monetary policy shocks on the macroeconomic variables of interest.

\subsection{Bayesian Estimation} \label{sec:Bayesian}

The details of the Bayesian estimation algorithm are provided in the Online Supplement. In this section, I briefly outline the main steps and emphasize the computational tractability of the proposed framework.

Estimation proceeds via a Gibbs sampler that iteratively samples from the complete conditional posterior distributions of each block of parameters. Despite incorporating several important extensions relative to \cite{Korobilis2022} --- namely, the inclusion of high-frequency instruments, time-varying loadings and volatilities, and fat-tailed idiosyncratic shocks --- all conditional posterior distributions remain available in closed form. This feature ensures that estimation remains computationally efficient even in very high-dimensional settings.

The key steps of the Gibbs sampler are as follows:
\begin{enumerate}
    \item \textbf{VAR coefficients $\bm \Phi$:} The autoregressive parameters are sampled equation-by-equation under a hierarchical Horseshoe prior that induces shrinkage and regularizes estimation in large VARs.
    \item \textbf{Factor loadings $\bm \Gamma$ and $\bm \Lambda$:} The factor loading matrices for the instruments and macro variables are sampled, respecting any imposed sign restrictions to facilitate structural interpretation. If certain loadings are time-varying, allowing exploration of time-varying IRFs, then a Horseshoe prior is used to shrink the expanded parameter space.
    \item \textbf{Latent factors $\bm f_{t}$:} The structural shocks (latent factors) are drawn conditional on the current parameters and data, ensuring orthogonality and unit variance normalization.
    \item \textbf{Idiosyncratic variances $\bm W$ and $\bm \Sigma$:} The variances of the idiosyncratic disturbances are sampled using standard inverse-gamma posteriors.
    \item \textbf{Time-varying loadings $\bm \Lambda_{t}$ and volatilities $\bm \Sigma_t$:} When time variation is allowed, the loadings and log-volatilities are modeled as random walks, and their entire trajectories are sampled using a stacked regression representation that avoids sequential Kalman filtering \citep{Korobilis2021}.
    \item \textbf{Degrees of freedom $\bm v$:} When allowing for leptokurtic errors, the degrees of freedom parameters governing the Student-t distributions of idiosyncratic shocks can be sampled using a standard Metropolis-within-Gibbs step \citep{Geweke1993}.
\end{enumerate}

By carefully designing the model structure and prior specification, the estimation algorithm avoids the need for computationally intensive steps. This makes the proposed framework highly scalable and well-suited for applications involving a large number of macroeconomic variables and predictors. In the empirical application, which involves a large VAR model with 30 variables (including the two instruments) and $p=2$ lags, it takes approximately two hours to generate 100,000 posterior draws using a desktop computer equipped with an Intel Core i9 12900K processor and 32 GB of RAM.

\section{Data}
The US macroeconomic dataset used in this paper spans January 1995 to May 2024 at a monthly frequency. Data transformations are selected to induce stationarity and align with standard VAR practice: levels for rate variables and indexes (interest rates, uncertainty, sentiment, financial conditions); first log differences for equity indices, exchange rates, and oil prices; and annual growth rates for output, inflation, and consumer spending. The use of annual growth rates reflects the smoother dynamics of these series, which helps produce clearer IRFs given that the data are not pre-filtered for outliers.

\begin{figure}
\centering
\includegraphics[width=0.7\textwidth,trim={5cm, 0cm, 5cm, 2cm}]{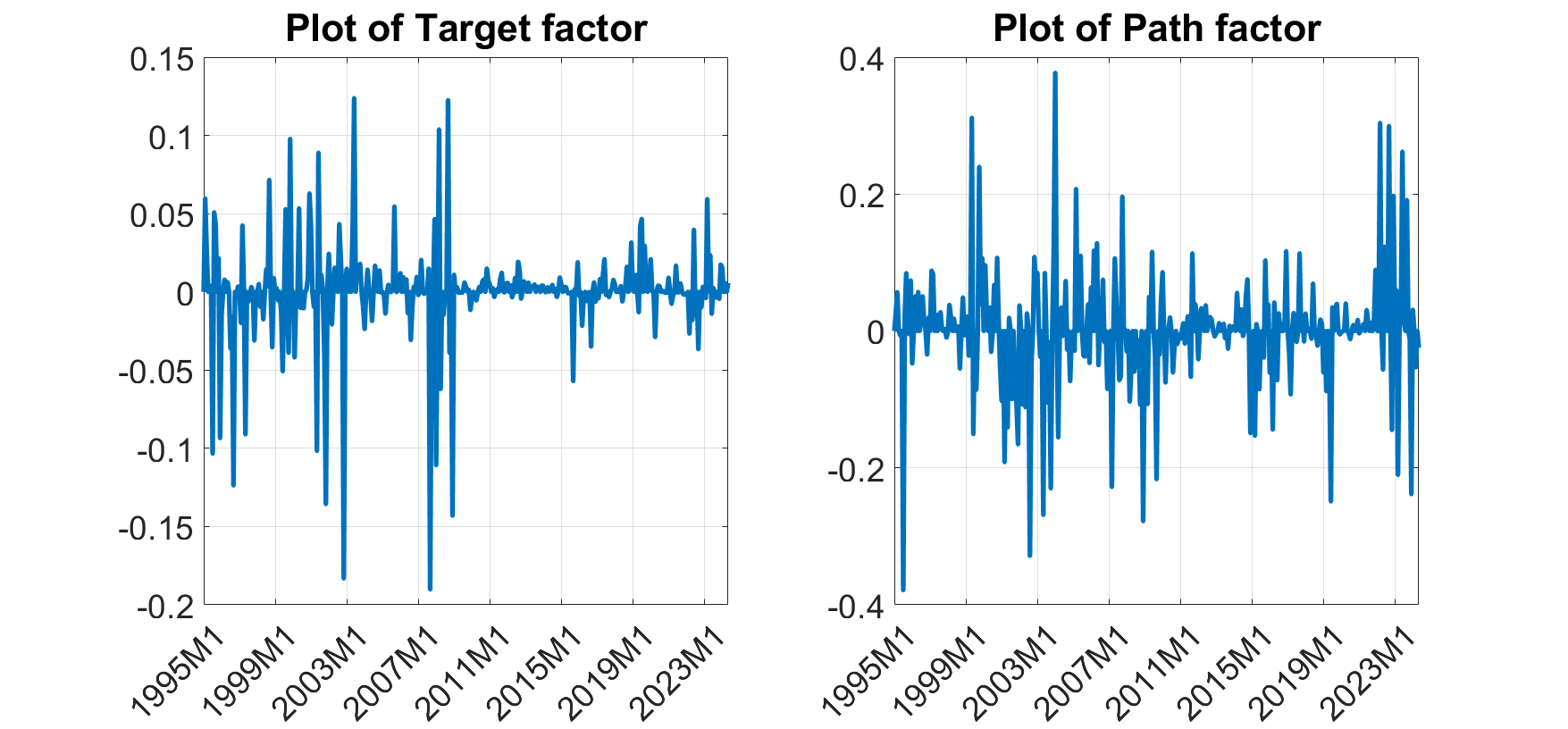}
\caption{Plots of target and path factors for the monthly sample 1995M1 to 2024M5.} \label{fig:surprises}
\end{figure}

The data fall into four broad categories: high-frequency monetary policy surprises, core aggregate macroeconomic variables, disaggregated CPI components, and predictor variables:
\paragraph{High-Frequency Monetary Policy Surprises} I follow \cite{Gurkaynaketal2005} in constructing two orthogonal factors from intradaily Treasury futures around FOMC announcements.
\begin{itemize}
\item Target factor: captures unexpected changes in the current federal funds rate target.
\item Path factor: measures surprises in the anticipated future path of policy (e.g. forward guidance).
\end{itemize}
Monthly aggregates of these high-frequency surprises are drawn from \cite{Acosta2024}, who constructs them by measuring 30-minute windows around scheduled FOMC announcements. In months when no announcements are available, the values of the surprises are set to zero, denoting the absence of information. \autoref{fig:surprises} plots these two series.

In the online supplement, I further assess robustness by re-estimating the VAR using the directly observed surprise series of \cite{Jarocinski2024}, imposing the exact sign restriction identification scheme of \cite{JarocinskiKaradi2020}; these data are publicly available on the author’s website.
\paragraph{Core Aggregate Macroeconomic Variables}
Our core VAR includes seven U.S. aggregates: real gross domestic product (GDP), personal consumption expenditures (PCE) inflation, the federal funds rate, 1- and 10-year Treasury yields, real M2 money and the Standard and Poor’s 500 (S\&P500) stock market index. For robustness, in the online supplement I consider several other monthly proxies of output and price inflation, such as total industrial production (IP) and core consumer price index (CPI).
\paragraph{Disaggregated CPI Components}
To study heterogeneous effects across consumption categories, I incorporate eight CPI subindices: Apparel; Education; Food; Other Goods; Housing; Medical Care; Recreation; and Transportation. These series are transformed into annual percentage changes to ensure comparability with aggregate inflation measures.
\paragraph{Other macro and predictor variables}
In addition to the seven core macro aggregates, the large VAR is augmented with a broad set of other macro and predictor series.
\bigskip

\autoref{tab:Data_table} details of all variables that are used in the remainder of this section. When a variable is originally available at a higher frequency (daily or weekly), monthly values are constructed by averaging all high frequency values over the month. All variables are downloaded from their sources seasonally adjusted, where this is relevant. The data were collected on February 12, 2025. Column Tcode shows short codes for applying stationarity transformations to the time series data, prior to estimating the VAR. The codes are as follows, 1: level, 2: first differences, 4: log-level, 5: first log differences (month-on-month growth rate), 7: annual differences (annual growth rate).

{\scriptsize
\begin{longtable}{p{0.03\textwidth}p{0.15\textwidth}p{0.40\textwidth}p{0.03\textwidth}p{0.3\textwidth}}
\caption{U.S. macro/financial variables, sources and definition} \\ \hline   \label{tab:Data_table}
No & Mnemonic & Description & Tcode & Source \\ \hline\hline
\multicolumn{4}{c}{\underline{\textsc{Proxies}}} \\
1 & Target  & Target factor of \cite{Gurkaynaketal2005} & 1 & \cite{Acosta2024}$^{1}$ \\
2 & Path  & Path factor of \cite{Gurkaynaketal2005} & 1 & \cite{Acosta2024}$^{1}$ \\
& \multicolumn{4}{c}{\underline{\textsc{Core macro variables whose response is sign-restricted}}} \\
3 & RGDP & Monthly real GDP & 7 & S\&P$^{2}$  \\
4 & PCE & Personal consumption expenditure deflator & 7 & FRED$^{3}$ \\
5 & SHADOW & Wu-Xia Shadow Federal Funds Rate & 1 & FRB$^{13}$ \\
6 & GS1 & Market Yield on U.S. Treasury Securities at 1-Year & 1 & FRED$^{3}$ \\
7 & GS10 & Market Yield on U.S. Treasury Securities at 10-Year & 1 & FRED$^{3}$ \\
8 & M2REAL & Real M2 Money Stock & 5 & FRED$^{3}$ \\
9 & SP500 & S\&P 500 price index & 5 & investing.com$^4$\\
& \multicolumn{4}{c}{\underline{\textsc{Disaggregate CPI variables used in the large-scale VAR}}} \\
10 & CPIA & CPI: Apparel & 7 & BLS$^{5}$ \\
11 & CPIE & CPI: Education & 7 & BLS$^{5}$ \\
12 & CPIF & CPI: Food & 7 & BLS$^{5}$ \\
13 & CPIO & CPI: Other goods & 7 & BLS$^{5}$ \\
14 & CPIH & CPI: Housing & 7 & BLS$^{5}$ \\
15 & CPIM & CPI: Medical care & 7 & BLS$^{5}$ \\
16 & CPIR & CPI: Recreation & 7 & BLS$^{5}$ \\
17 & CPIT & CPI: Transportation & 7 & BLS$^{5}$ \\
& \multicolumn{4}{c}{\underline{\textsc{Additional predictor variables used in the large-scale VAR}}} \\
18 & UNRATE & Unemployment Rate & 1 & FRED$^{3}$ \\
19 & HOUST & New Privately-Owned Housing Units Started & 4 & FRED$^{3}$ \\
20 & RCONS & Real Personal Consumption Expenditures & 5 & FRED$^{3}$ \\
21 & TWEXAFEGSMTHx & Nominal Major Currencies U.S. Dollar Index & 5 & FRED$^{3}$ \\
22 & MCOILBRENTEU & Crude Oil Prices: Brent - Europe & 5 & FRED$^{3}$ \\
23 & USEPU & Global economic policy uncertainty & 1 &  EPU website$^{6}$\\
24 & GPR & Geopolitical risk index & 1 & \cite{CaldaraIacoviello2022}$^{7}$ \\
25 & EBP & Excess bond premium & 1 & FRB$^{8}$ \\
26 & JLNF12 & Financial uncertainty & 1 & \cite{jurado2015measuring}$^{9}$ \\
27 & UMSCENT & University of Michigan Consumer Sentiment & 1 & FRED$^{3}$ \\
28 & VIXCLS & CBOE Volatility Index: VIX & 1 & FRED$^{3}$ \\
29 & MPU & Monetary Policy Uncertainty & 1 & FRB$^{10}$ \\
30 & GFC & Global Financial Cycle & 1 & \cite{miranda-agrippino2020us}$^{11}$ \\ 
& \multicolumn{4}{c}{\underline{\textsc{Alternative measures}}} \\
1$^{\star}$/2$^{\star}$ & MP1 & Fed funds futures (surprises) & 1 & \cite{Jarocinski2024}$^{12}$ \\ 
1$^{\star}$/2$^{\star}$ & TFUT10 &   10-year U.S. Treasury Note futures (surprises) & 1 & \cite{Jarocinski2024}$^{12}$ \\
1$^{\star}$/2$^{\star}$ & SP500FUT & S\&P futures (surprises)  & 1 & \cite{Jarocinski2024}$^{12}$ \\
3$^{\star}$ & BBKMGDP & Brave-Butters-Kelley Real Gross Domestic Product & 1 & FRED$^{3}$ \\
3$^{\star}$ & INDPRO & INDPRO & 7 & FRED$^{3}$ \\
4$^{\star}$ & CPILFESL & CPI: All Items Less Food and Energy & 7 & FRED$^{3}$ \\
5$^{\star}$ &  FFR & Federal Funds Effective Rate & 1 & FRED$^{3}$ \\
\hline
\multicolumn{5}{l}{{\tiny $^1$Author's webpage \href{https://www.acostamiguel.com/replication/MPshocksAcosta.xlsx}{https://www.acostamiguel.com/replication/MPshocksAcosta.xlsx}} }\\
\multicolumn{5}{l}{ {\tiny  $^{2}$S\&P Market Intelligence \href{https://www.spglobal.com/marketintelligence/en/mi/products/us-monthly-gdp-index.html}{https://www.spglobal.com/marketintelligence/en/mi/products/us-monthly-gdp-index.html}} } \\
\multicolumn{5}{l}{ {\tiny $^{3}$Federal Reserve Economic Data \href{https://fred.stlouisfed.org/}{https://fred.stlouisfed.org/}} } \\
\multicolumn{5}{l}{ {\tiny $^{4}$Investing.com \href{https://www.investing.com/indices/us-spx-500-historical-data}{https://www.investing.com/indices/us-spx-500-historical-data} } } \\
\multicolumn{5}{l}{ {\tiny $^{5}$Bureau of Labor Statistics \href{https://www.bls.gov/cpi/data.htm}{https://www.bls.gov/cpi/data.htm} } } \\
\multicolumn{5}{l}{ {\tiny $^{6}$\href{https://www.policyuncertainty.com/global\_monthly.html}{https://www.policyuncertainty.com/global\_monthly.html} } } \\
\multicolumn{5}{l}{{\tiny $^{7}$Author's website \href{https://www.matteoiacoviello.com/gpr.htm}{https://www.matteoiacoviello.com/gpr.htm}} }  \\
\multicolumn{5}{l}{ {\tiny $^{8}$Federal Reserve Board \href{https://www.federalreserve.gov/econres/notes/feds-notes/ebp\_csv.csv}{https://www.federalreserve.gov/econres/notes/feds-notes/ebp\_csv.csv}} } \\
\multicolumn{5}{l}{ {\tiny $^{9}$Authors' website \href{https://www.sydneyludvigson.com/macro-and-financial-uncertainty-indexes}{https://www.sydneyludvigson.com/macro-and-financial-uncertainty-indexes}} }  \\
\multicolumn{5}{l}{ {\tiny $^{10}$Federal Reserve Board \href{https://www.frbsf.org/research-and-insights/data-and-indicators/market-based-monetary-policy-uncertainty/}{https://www.frbsf.org/research-and-insights/data-and-indicators/market-based-monetary-policy-uncertainty/}} } \\
\multicolumn{5}{l}{ {\tiny $^{11}$Author's website \href{https://silviamirandaagrippino.com/code-data}{https://silviamirandaagrippino.com/code-data}} } \\
\multicolumn{5}{l}{ {\tiny $^{12}$Author's website \href{https://github.com/marekjarocinski/tshocks\_fomc\_update}{https://github.com/marekjarocinski/tshocks\_fomc\_update}} } \\
\multicolumn{5}{l}{ {\tiny $^{13}$Federal Reserve Bank of Atlanta \href{https://www.atlantafed.org/cqer/research/wu-xia-shadow-federal-funds-rate}{https://www.atlantafed.org/cqer/research/wu-xia-shadow-federal-funds-rate}} }  
\end{longtable}
}

\section{Empirics}
The empirical section is structured in two parts. I first estimate a small VAR that includes only the external instruments (monetary policy surprises) and a core set of macroeconomic variables whose responses to policy are assumed to be stable over time and are identified using sign restrictions. Therefore, the model consists of variables 1-9 in \autoref{tab:Data_table}. The VAR features stochastic volatility and student-t idiosyncratic errors, as changes in volatility and outliers are present in all macro data. This small-scale VAR serves as a baseline model and illustrates the effectiveness of the identification strategy that combines sign and instrument-based restrictions. The results show that conventional monetary policy shocks can be identified robustly across a wide range of modeling choices, including the definition of instruments (target and path factors versus observed surprises), the choice of macroeconomic indicators (such as different measures of output, prices, or interest rates), and the specification of the VAR itself (including lag length, instrument exogeneity assumptions, and orthogonality conditions).

In the final part of the section, I expand the VAR by incorporating a richer set of variables, including disaggregated consumer price indices and forward-looking predictor variables. This model corresponds to the first 30 variables in \autoref{tab:Data_table}. The primary objective of this extension is to uncover how different components of the CPI responded to monetary policy after 2021, without imposing strong structural assumptions on the behavior of these disaggregated series. To this end, I retain the sign restrictions on the core macroeconomic variables to ensure identification of structural shocks in line with theory, while leaving the responses of the additional variables unrestricted and free to vary over time. These unrestricted impulse responses are entirely shaped by the data and are regularized through the use of shrinkage. This design allows the model to isolate structural shocks using theory-consistent restrictions on aggregate variables, and to trace out their effects on disaggregated prices in a flexible and data-driven way.

\subsection{Monetary policy: small benchmark model}\label{sec:smallVAR} 

The first VAR is a small-scale benchmark model containing the two high-frequency monetary policy surprise factors (the ``target'' and ``path'' factors) alongside a core set of U.S. macroeconomic variables. The core endogenous variables include seven key aggregates observed at a monthly frequency – real economic activity (output), inflation, the policy rate\footnote{In the benchmark specification I use the \citep{WuXia2016} shadow rate.} and short-term interest rates, a long-term interest rate, a monetary aggregate, and stock prices. I set the VAR lag order of this nine-variable VAR to $p=6$ (months) in the baseline and I allow for stochastic volatility and student-$t$ errors to mitigate the influence of outliers and changes in volatility. However, for the sake of solidifying identification, these seven core macroeconomic variables have a time-invariant response to monetary policy shocks. In terms of the VAR formulations already introduced, this assumption corresponds to the model in equation \eqref{largeMPVAR2} with the restriction $\bm \Lambda_{t} = \bm \Lambda$. This small VAR serves to illustrate conventional monetary transmission in a controlled setting before moving to a higher-dimensional model.

\paragraph{Identification strategy.} I identify two structural monetary policy shocks in this VAR: a \emph{Target shock}, corresponding to an unanticipated change in the current policy rate (conventional policy shock), and a \emph{Path shock}, corresponding to news about future policy. The proposed identification strategy combines high-frequency external instruments with sign and zero restrictions on impact responses of select variables. High-frequency surprises around FOMC announcements are powerful proxies for policy shocks because they are measured in a narrow window that filters out most endogenous market responses. However, using these surprises alone to identify shocks can be problematic if they contain central-bank information effects or other confounding influences.

\begin{figure}[H]
\centering
\includegraphics[width=\textwidth,trim={5cm, 1cm, 5cm, 1cm}]{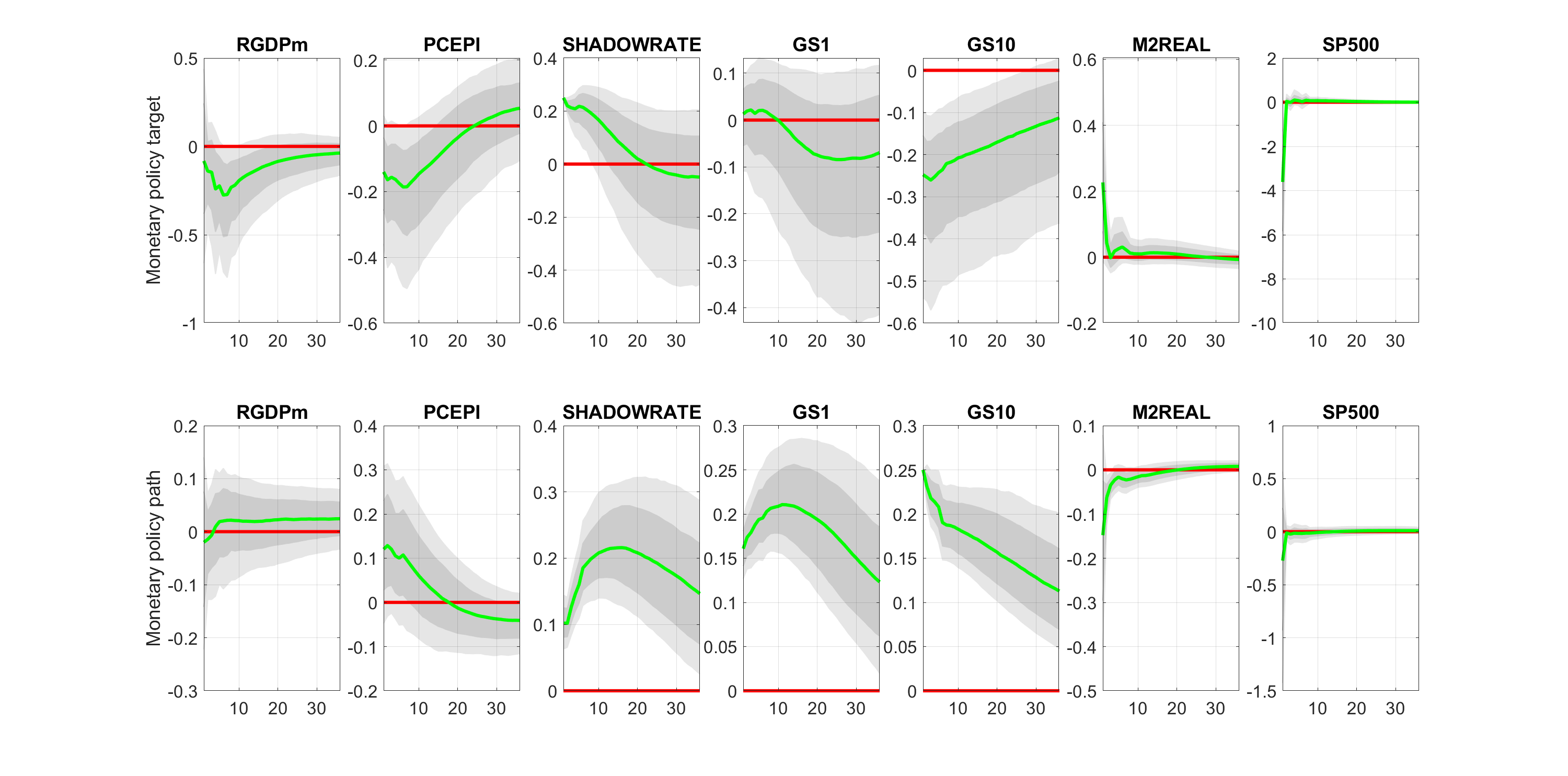}   
\caption{Responses of macro variables to Target (top row) and Path (bottom row) shocks in benchmark VAR without sign restrictions on macro variables. Solid green lines show the posterior median response, while dark grey and light grey bands represent the 68\% and 90\% posterior credible intervals, respectively; these summarize uncertainty across the posterior distribution of the structural impulse responses, reflecting both estimation error and parameter uncertainty.}\label{IRFnosign}
\end{figure}

Indeed, an initial attempt using only the two surprise factors to identify the shocks (imposing that the Target shock is the only driver of the target factor and the Path shock the only driver of the path factor, with no other restrictions) produced some counterintuitive results. As shown in \autoref{IRFnosign}, a putative contractionary Target shock (tightening of the policy rate) did not yield a clearly negative effect on output or inflation – the posterior median responses of real GDP and PCE inflation were slightly negative but not significant – and it even led to a decline in the 10-year yield and a rise in real M2, contrary to typical theory predictions. Similarly, the identified Path shock had little effect on real activity and was associated with a slight rise in prices on impact. These anomalies echo concerns in the literature that high-frequency policy surprises may be contaminated by ``information shocks'' (when agents infer positive news about the economy from a rate hike) or may suffer from weak instrument problems, leading to implausible impulse responses. Recent studies have emphasized the need to disentangle pure policy shocks from central bank information components \citep[e.g.,][]{Jarocinski2024}, and the findings here reinforce that message.

\begin{table}
\centering
\begin{tabular}{ccccc} \hline
 & \multicolumn{2}{c}{Identified shocks} &  \multicolumn{2}{c}{Residual shocks}  \\
Endogenous vars  &  Target shock & Path shock & Residual shock 1 & Residual shock 2 \\  \hline \hline
$\begin{array}{c}
Target \\
Path \\
----- \\
real GDP \\
PCE inf  \\
FFR \\
GS1 \\
GS10 \\
M2REAL \\
SP500
\end{array}$  &
$\begin{array}{c}
+ \\
0 \\
----- \\
- \\
-  \\
+ \\
+ \\
0 \\
-  \\
-
\end{array}$
&
$\begin{array}{c}
0 \\
+ \\
----- \\
NA \\
0 \\
0 \\
NA \\
+ \\
NA \\
NA
\end{array}$
&
$\begin{array}{c}
0 \\
0 \\
----- \\
NA \\
NA \\
NA \\
NA \\
NA \\
NA \\
NA
\end{array}$
&
$\begin{array}{c}
0 \\
0 \\
----- \\
NA \\
NA \\
NA \\
NA \\
NA \\
NA \\
NA
\end{array}$  \\ \hline
\end{tabular}
\caption{Sign restrictions imposed to core macro variables, which correspond to sign restrictions to the matrix $\bm \Lambda$. Matrix $\bm \Lambda$ here is constant, not time-varying.} \label{tab:identificationscheme}
\end{table}

To pin down a conventional monetary tightening shock more robustly, I impose additional sign and zero restrictions on impact responses of the core macro variables, guided by economic theory. Table~\ref{tab:identificationscheme} summarizes the identification scheme. A Target shock is defined as an unanticipated rate hike that requires: (i) the target factor (current FFR surprise) to react positively, (ii) the path factor (expected future rate surprise) to be zero on impact, and (iii) core macro variables move in line with textbook predictions. With regards to the last point, output and the price level \emph{fall} on impact signifying a contraction in demand and disinflationary pressure, the policy rate and short-term interest rates rise, real money supply falls due to tightening liquidity and credit conditions, and equity prices fall reflecting higher discount rates and weaker expected earnings. I leave the response of the 10-year yield unrestricted; prior intuition would suggest it likely rises or remains flat, as a monetary tightening can either raise long-term rates via expectations of future short rates or, in some cases, lower them if it strongly depresses future growth/inflation expectations. For the Path shock, I impose: (i) zero impact on the current policy rate (target factor) and a positive impact on the path factor by definition, and (ii) no contemporaneous change in output and zero or muted impact on inflation.\footnote{I do not force inflation down for a path shock, allowing the data to determine if this shock acts like news of future policy.} A positive Path shock raises the 10-year yield on impact, since this shock should shift expectations of the rate trajectory. Other variables’ responses to the Path shock are left largely unrestricted. These hybrid identification restrictions – combining external instrument relevance with sign constraints on macroeconomic aggregates – ensure that the Target shock recovered corresponds to a standard contractionary monetary policy shock (stripped of information effects), while the Path shock is identified as a distinct structural disturbance capturing revisions to expected future policy.\footnote{On top of these restrictions, in \autoref{tab:identificationscheme} I specify two residual shocks that are not identified. This is due to the fact that in the proposed VAR model the number of shocks assumed not only affects identification, but also estimation accuracy. The number of shocks corresponds to the number of the estimated factors $\bm f_{t}$, which in turn affect how precisely-estimated is the ``true'' VAR covariance matrix $\bm \Omega$. Note that these additional residual factors need not be identified; the common component $\bm \Lambda \bm f_{t}$ of these two factors is identified and this is sufficient for estimation using MCMC.}

\begin{figure}[H]
\centering
\includegraphics[width=\textwidth,trim={5cm, 1cm, 5cm, 1cm}]{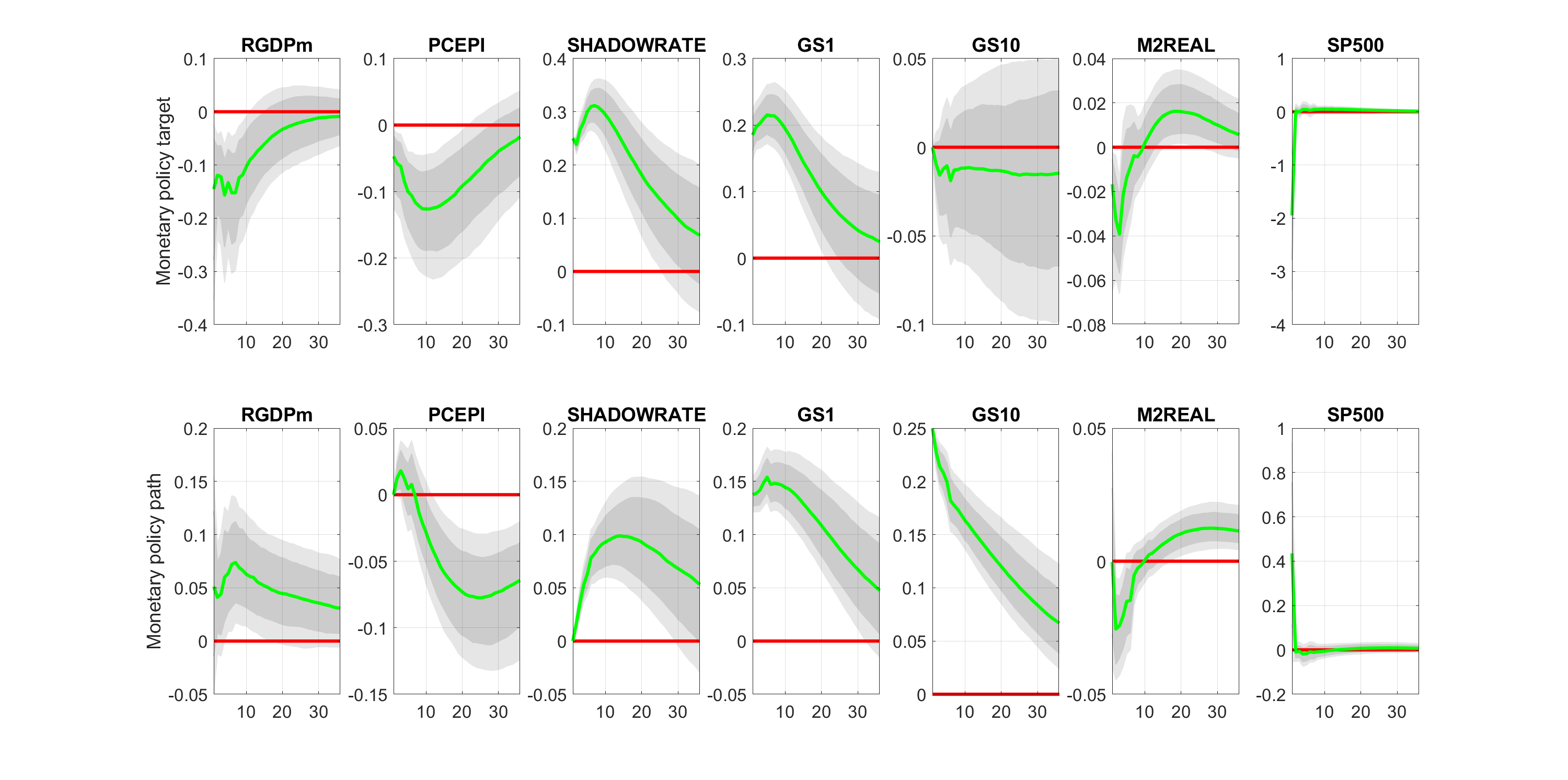}    
\caption{Responses of macro variables to Target (top row) and Path (bottom row) shocks in benchmark VAR with sign restrictions on macro variables that follow \autoref{tab:identificationscheme}. Solid green lines show the posterior median response, while dark grey and light grey bands represent the 68\% and 90\% posterior credible intervals, respectively; these summarize uncertainty across the posterior distribution of the structural impulse responses, reflecting both estimation error and parameter uncertainty.}\label{IRFsign}
\end{figure}

Figure \ref{IRFsign} presents the impulse response functions (IRFs) of the core seven macro variables to a contractionary monetary policy shock under the preferred identification (combining the surprise instruments with sign restrictions). The responses are economically sensible and in line with conventional monetary transmission mechanisms and previous empirical evidence. Overall, the small VAR results demonstrate classic monetary transmission channels at work. A true contractionary policy shock (Target factor) leads to: higher short-term interest rates, a broad-based tightening of financial conditions (less money, lower equity values, likely higher credit spreads), a decline in aggregate demand (output falls) and a moderation in inflation. These outcomes accord with the broad consensus from decades of VAR studies and with standard New Keynesian theory (monetary tightening moves the economy along the aggregate demand curve, and the Phillips curve translates the induced slack into lower inflation). Importantly, these effects are estimated on data that extend through the recent high-inflation episode of 2021–2022, suggesting that even in that atypical period, the fundamental channels remained operative. The Path shock, meanwhile, illustrates the more complex interplay of policy signaling and expectations. While it does not strongly contract the economy on impact, its existence highlights that not all Fed actions map neatly onto ``surprise tightenings'' – some reflect information updates. 

In the subsequent analysis with a larger VAR this identification of two separate monetary shocks is maintained. Nevertheless, the focus there is on the conventional Target shock as the driver of systematic contractionary policy actions, while acknowledging the Path shock as an important but less clear-cut disturbance. Before moving to the large-scale model, I verify that the small VAR results are robust to various alternative specifications. In the Online Supplement I report detailed robustness checks, but I summarize them briefly here. First, increasing the lag order to $p=12$ or using tighter/slacker priors yields very similar IRFs, indicating that the baseline lag length (6) is sufficient to capture the dynamics. Second, I considered alternative identifying assumptions suggested by \citet{JarocinskiKaradi2020}: (a) restricting the lagged influence of the surprises on macro variables ($\Phi_{ym}=\mathbf{0}$, so that the high-frequency shocks only enter contemporaneously and not in lagged form), and (b) conversely restricting the feedback of macro shocks on the surprise series ($\Phi_{mm}=\mathbf{0}$). Imposing these did not materially change the identified IRFs of interest – the output and inflation responses to a Target shock remained significantly negative and of similar magnitude. Third, replacing the \citep{WuXia2016} shadow rate with the effective Federal funds rate does not alter the qualitative pattern of responses of output and inflation, although the response of the 10-year rate becomes significantly negative. Finally, allowing the Target and Path shocks to be correlated (relaxing the orthogonality assumption) did not appreciably affect the impulse responses, though for our baseline I keep them orthogonal for a cleaner structural interpretation. All these checks bolster the confidence that the small VAR captures a genuine monetary tightening shock and its transmission in a theoretically consistent manner. In what follows, I build on this foundation to examine a richer array of variables and potential time variation in responses.

\subsection{Empirical results: large VAR with time-varying disaggregated CPI responses}\label{sec:largeVAR} 

I now turn to a large-scale Bayesian VAR that incorporates a much broader information set, including disaggregated price indices and financial market indicators. The primary goal of this extension is to study how a monetary policy shock propagates across different sectors and over time – in particular, how various components of the consumer price index (CPI) respond (heterogeneously) to policy tightening, and how financial conditions and expectations evolve alongside. By expanding the variable set, we also allow the possibility for the effects of monetary policy to change over the sample, since certain relationships may not be constant in the face of structural changes (e.g., the shift from a low-inflation regime to the high-inflation environment after 2021). The large VAR includes 30 monthly variables: the same two policy surprise factors and seven core macro variables from the small VAR, plus 13 additional macro-financial predictor series and the eight disaggregated CPI sub-indices. The benchmark case uses $p=2$ lags of all variables, as this choice yields more stable and interpretable impulse response functions. The Online Supplement shows that results are qualitatively unchanged when $p=6$.

The identification of the monetary policy shocks in the large VAR follows the \emph{same scheme as in the small VAR}: I impose the sign and zero restrictions on the impact responses of the core macro block (real activity down, inflation down for the Target shock; policy rate up, etc., as detailed earlier in Table~\ref{tab:identificationscheme}), and similarly use the high-frequency target/path instruments. No additional sign restrictions are imposed on the new variables – instead, I ``let the data speak'' for how those variables respond. This design – identify shocks with a theoretically grounded core subset, but leave the rest unrestricted – is akin to a factor-augmented VAR or a large panel SVAR analysis. It ensures that the monetary policy shock that is recovered in the large model is directly comparable to the one in the small model, while leveraging the larger information set to observe richer dynamics and potential time variation. 

\begin{figure}[H]
\centering
\includegraphics[width=\textwidth,trim={5cm, 1cm, 5cm, 1cm}]{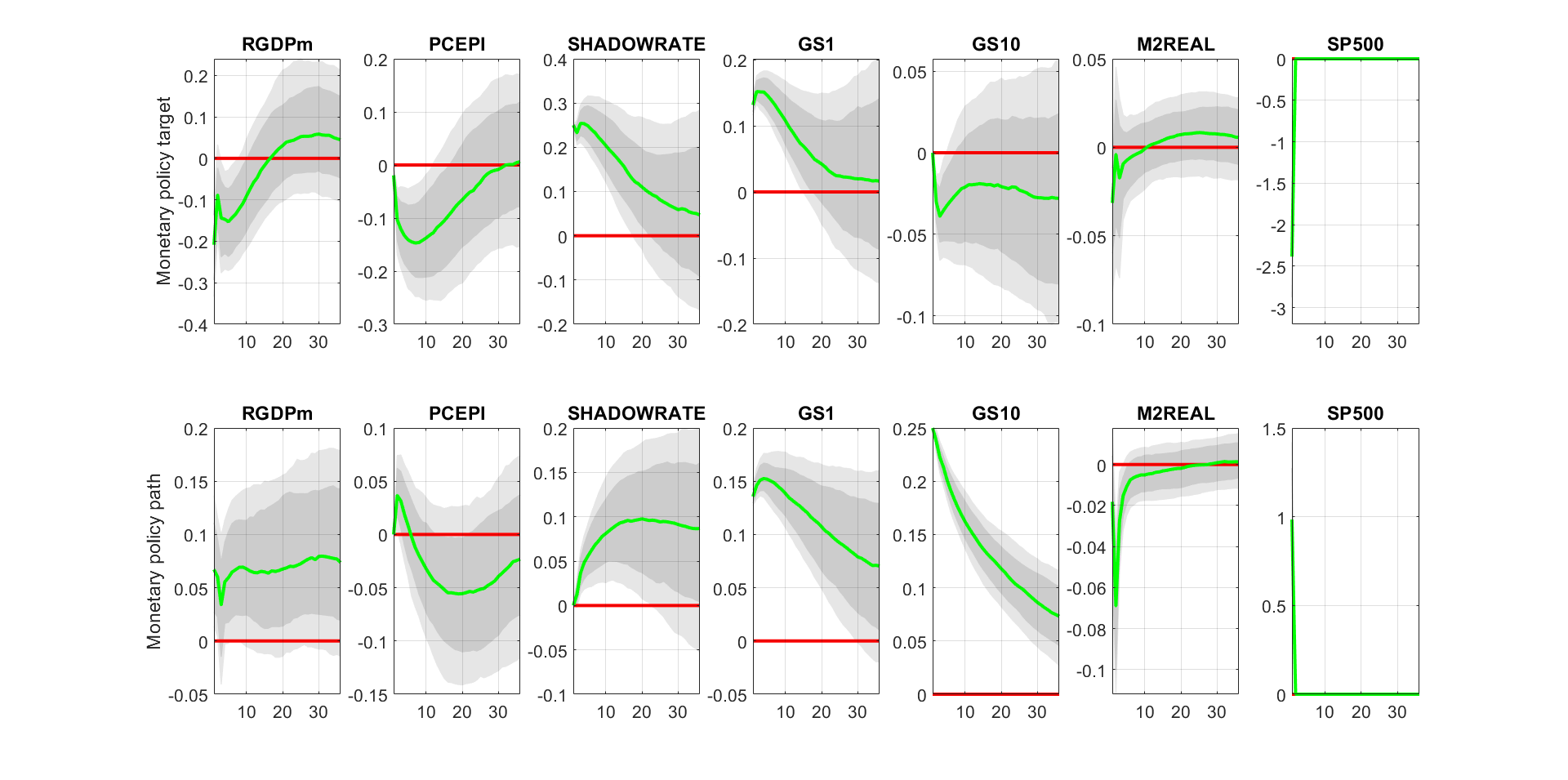} 
\caption{Responses of core macro variables to Target and Path shocks in the large VAR. $\bm \Lambda$ is constant for these variables, therefore, IRFs are fixed over the sample.}\label{IRFcore}
\end{figure}

\paragraph{Validating core responses in the large VAR.} I first verify that the introduction of many additional variables, and allowing for time-varying coefficients, does not distort the baseline monetary transmission mechanism. Figure~\ref{IRFcore} plots the impulse responses of the core seven macro variables to the Target and Path shocks as estimated in the large VAR. These core responses are restricted to be constant over time, hence the figure represents the common response for the entire sample. Reassuringly, the core IRFs are fairly similar to those from the small VAR (Figure~\ref{IRFsign}). A contractionary Target shock in the large VAR still produces a clear downturn in real activity, a gradual decline in inflation, an uptick in short-term interest rates, and tightening of money and equity markets, all with magnitudes and timing comparable to the small-model results. In fact, the peak output and inflation effects are nearly identical, indicating that the identification is stable and the additional variables have not introduced any ambiguity about the nature of the shock. This consistency is important: it implies that our shock remains a ``conventional'' monetary policy shock, and any new insights in the large VAR will come from observing the behavior of the additional variables rather than from redefining the shock itself. It also highlights that the Bayesian shrinkage and time-variation do not undermine the core inference; rather, the prior helps to integrate the extra information without losing the signal from the primary series. In summary, the large VAR’s core results pass a key sanity check giving confidence to proceed with analyzing the richer dynamics.

\begin{figure}[H]
\centering
\includegraphics[width=\textwidth,trim={5cm, 1cm, 5cm, 1cm}]{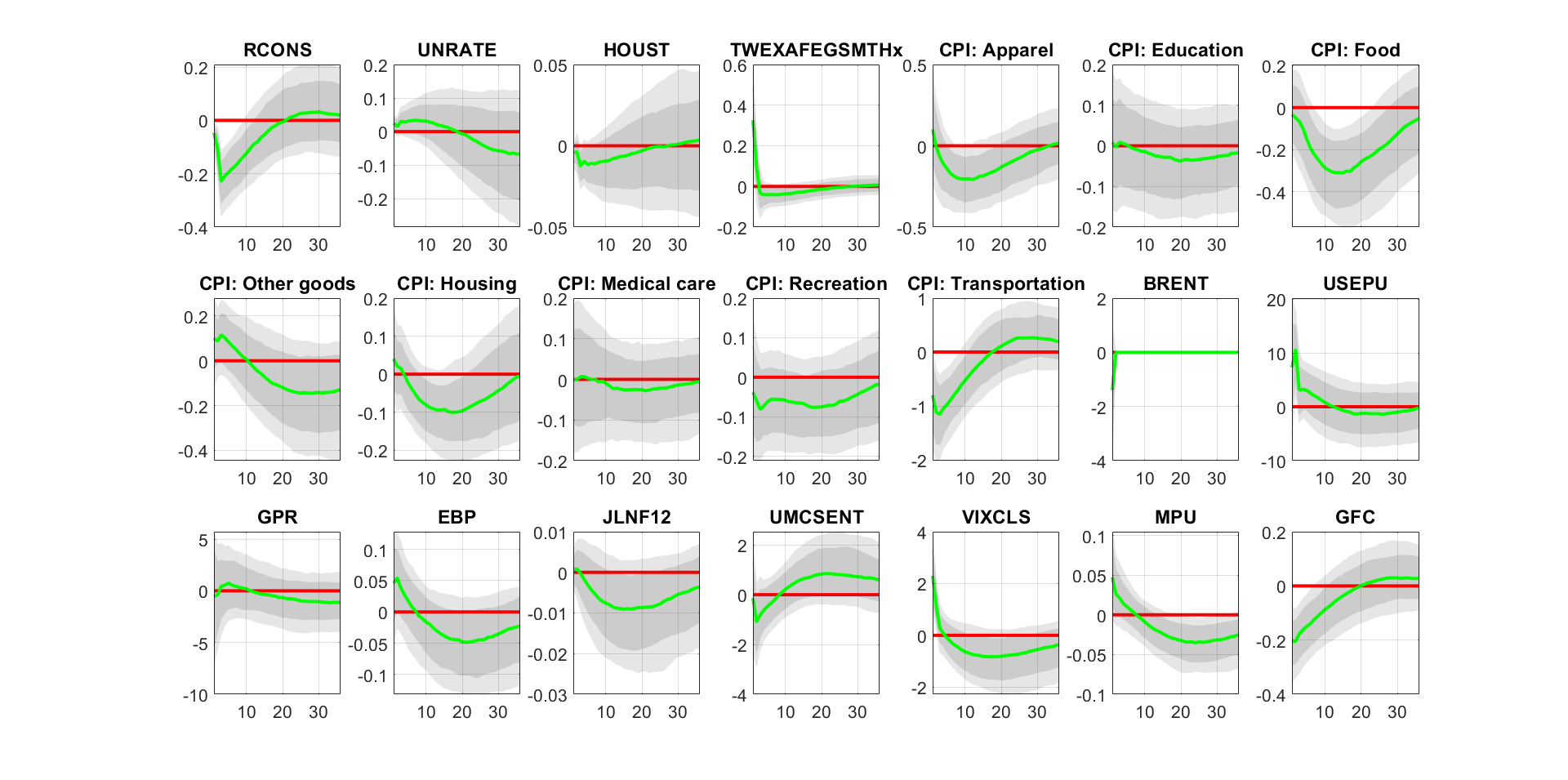}  
\caption{Responses of all other macro variables to Target shock in the large VAR. $\bm \Lambda$ is time-varying for these variables, therefore, plot gives responses at time $T$.}\label{IRFTargetother}
\end{figure}

\begin{figure}[H]
\centering
\includegraphics[width=\textwidth,trim={5cm, 1cm, 5cm, 1cm}]{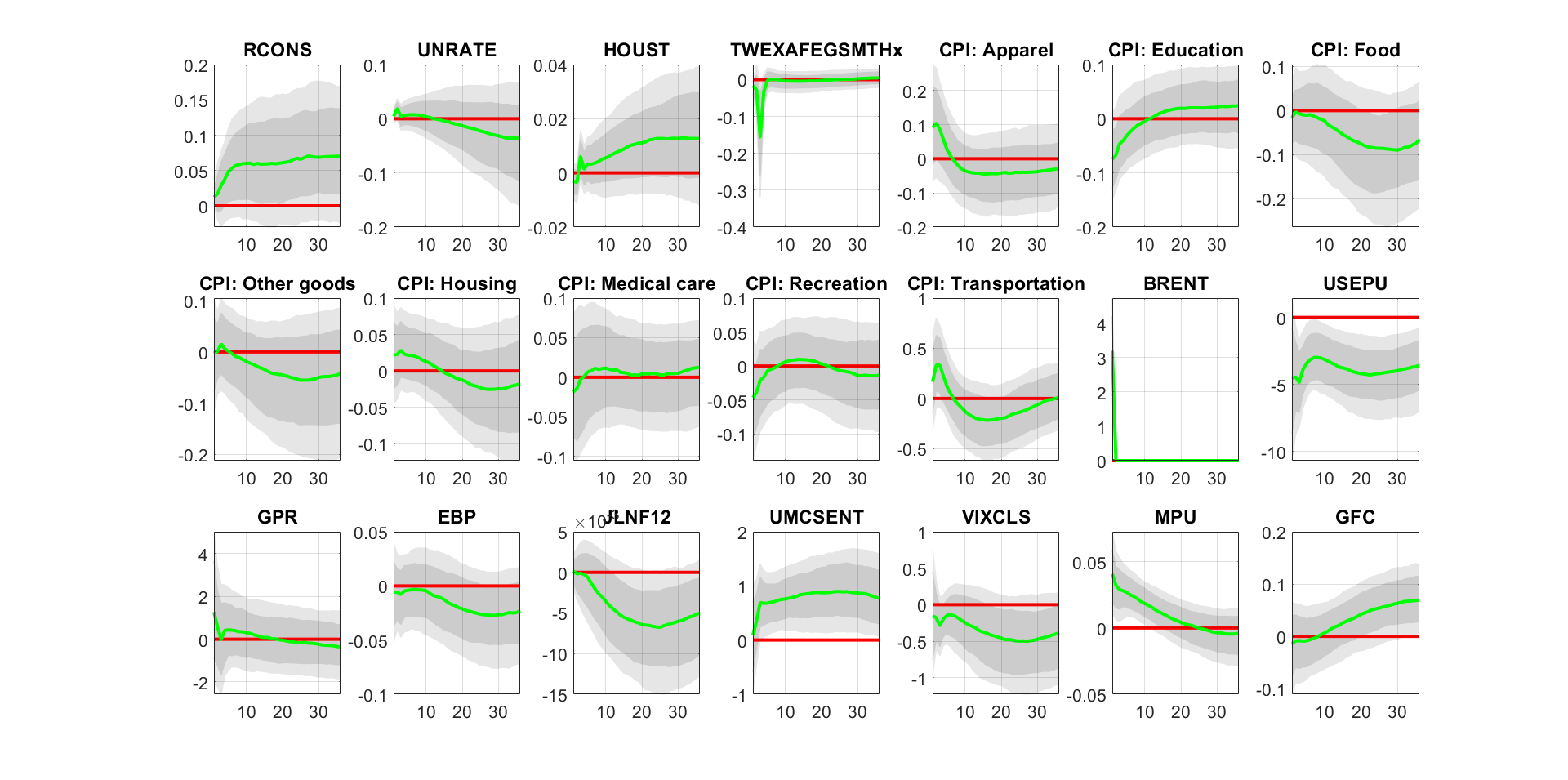} 
\caption{Responses of all other macro variables to Path shock in the large VAR. $\bm \Lambda$ is time-varying for these variables, therefore, plot gives responses at time $T$.} \label{IRFPathother}
\end{figure}

\paragraph{Effects on broader financial conditions and expectations.} Expanding the VAR to include more financial indicators allows us to paint a more detailed picture of how monetary policy tightening percolates through financial markets and expectations. Figure~\ref{IRFTargetother} shows the impulse responses at the end of sample (May 2024) of all the additional variables to a Target shock, while Figure~\ref{IRFPathother} does the same for a Path shock.\footnote{Note that time-varying impact responses to these variables are analyzed later.} 

I focus first on the Target shock (the conventional tightening). In short horizons the real variables have the correct sign: real consumption falls significantly, while unemployment rises and housing starts fall, although very marginally and insignificantly. A rate hike makes lenders and investors more risk-averse, leading them to demand higher premiums on risky debt, which in turn tightens credit supply to firms. The evidence in \autoref{IRFTargetother} confirms that the monetary shock triggers a cascade where financial conditions deteriorate significantly. The Excess Bond Premium (EBP) captures the risk-related component of credit spreads beyond expected defaults, and it typically rises after a contractionary monetary policy shock as investors become more risk-averse and demand higher compensation for holding corporate debt. Volatility (VIXCLS) and uncertainty measures (USEPU, MPU) also rise, reflecting increased trading volume and forward-looking uncertainty.

The responses to a Path shock (Figure~\ref{IRFPathother}, bottom panels) differ qualitatively, underscoring that this shock carries an informational element. Importantly, a Path shock does \emph{not} cause credit spreads to widen – in fact, if anything, the estimates suggest a slight \emph{narrowing} of corporate spreads immediately after a positive Path shock. At the same time, equity prices tend to increase on impact as the stock market often rallies in response to what appears to be a hawkish signal about the future. These financial reactions are consistent with a scenario in which the Fed’s guidance about future tightening is interpreted as a sign of confidence in economic strength. In other words, the Path shock behaves much like a positive ``news'' shock about the economy. This mirrors the \cite{JarocinskiKaradi2020} result that an information shock (good news) leads to higher stock prices and lower credit spreads even as rates rise. In terms of other effects, a Path shock results in a non-significant increase in real consumption and housing starts, a significant currency depreciation, and decrease of economic policy uncertainty and stock market volatility. Unsurprisingly, monetary policy uncertainty does not increase, rather it rises significantly to this central-bank induced shock.

In summary, the large VAR confirms that the two identified shocks have distinct financial market and expectations footprints: the Target shock is an all-around contractionary disturbance with immediate tightening of financial conditions, while the Path shock is more of an expectations shock that can even be interpreted as “good news” at the moment of impact, with its contractionary bite coming later.

\paragraph{Heterogeneous and time-varying inflation responses.} 
A central motivation for the large VAR is to examine how different components of inflation react to monetary policy shocks, and whether these reactions have changed in the recent high-inflation period.


\begin{figure}[H]
\centering
\includegraphics[width=\textwidth,trim={5cm, 1cm, 5cm, 1cm}]{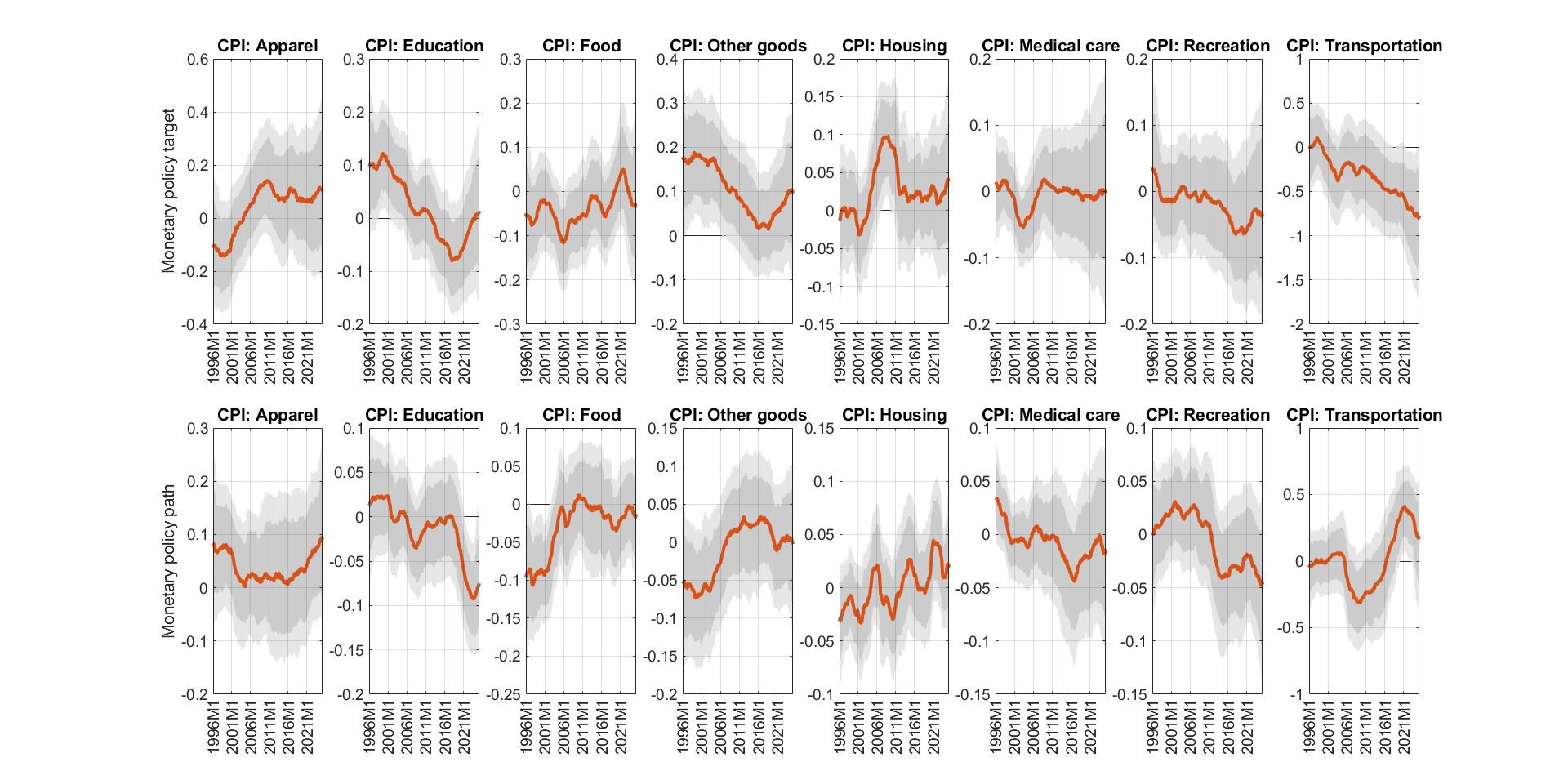}  
\caption{Impact responses of disaggregated CPI variables to Target and Path shocks.}\label{IRFimpactCPI}
\end{figure}

\begin{figure}[H]
\centering
\includegraphics[width=\textwidth,trim={5cm, 1cm, 5cm, 1cm}]{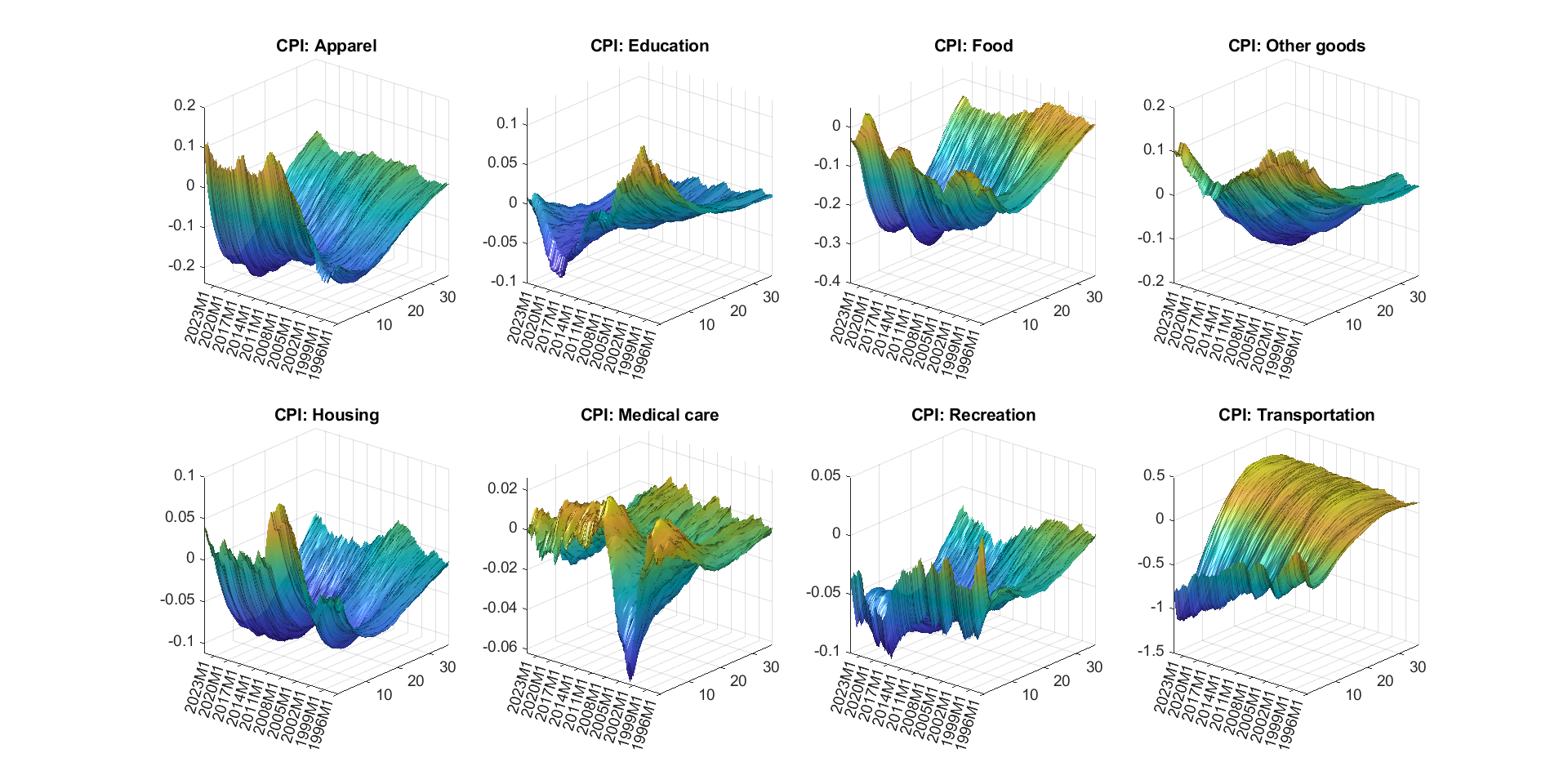}  
\caption{Responses over time and over horizons of disaggregated CPI variables to a Target shock.}\label{IRFTVPCPI}
\end{figure}

Figure~\ref{IRFimpactCPI} presents the \emph{impact} responses (at time $t=0$) of the eight disaggregated CPI categories to the Target and Path shocks, estimated at the end of our sample (i.e. for a shock occurring in May 2024). Figure~\ref{IRFTVPCPI} goes further, showing the evolution of the entire impulse response function of these CPI components to a Target shock over time (plotting the median responses at various horizons for shocks in different years). 

Several noteworthy patterns emerge. First, the impact responses in Figure~\ref{IRFimpactCPI} reveal considerable heterogeneity across sectors in the instantaneous effect of a monetary tightening. A contractionary Target shock immediately lowers the rate of inflation in some categories, while others show virtually no reaction on impact. Rent inflation (CPI: Housing), which is a notoriously sticky and inertia-driven component of total inflation, shows essentially no movement on impact. Landlords do not reprice rental contracts in direct response to a rate hike; instead, the effect of monetary policy on rents operates with a long lag, via cooling of the housing market and slower rent growth over time. I indeed find in the three-dimensional IRFs that the response of rent inflation to a Target shock, while negligible on impact, becomes slightly negative a year or two after the shock – indicating that monetary tightening eventually tempers rent increases, but only gradually. Core goods prices (e.g., furnishings, electronics, vehicles), which are more interest-sensitive because they often involve credit financing (auto loans, for instance) and have more flexible pricing than services, show a notable immediate decline. For instance, the price index for vehicles and durable goods drops on impact after the Target shock, reflecting a sudden pullback in demand; higher loan rates discourage car and appliance purchases, leading retailers and manufacturers to offer discounts or slow price increases to stimulate sales. Meanwhile, food prices, which can be volatile but also heavily influenced by supply (weather, global commodity markets), show a very mild impact response – perhaps a slight dip, but within credible bands that include zero. Food and beverages may not immediately respond to domestic monetary conditions, except insofar as general demand changes slowly feed through to lower markups in food retail.

Turning to the Path shock impact responses: these are generally much smaller in absolute magnitude (often not significantly different from zero) for all CPI components. This again underscores that an expected future tightening alone does not immediately move prices – likely because neither aggregate demand nor costs are materially affected on impact. If anything, we can see a hint that a positive Path shock might cause a slight \emph{uptick} in some prices like stock-sensitive goods or perhaps energy, consistent with the idea that it signals stronger future activity. However, these estimates are not very precise, and we should not ascribe too much economic significance to small positive bumps. The key takeaway is that unlike the Target shock, the Path shock does not uniformly push prices down at impact. Thus, from a policymaker’s perspective, an announced future hike (without current action) is not a substitute for a current hike in terms of immediately curbing inflation – its influence on prices will come later, mediated by expectations and subsequent activity changes. 

Perhaps the most interesting insights come from the time variation in the Target shock responses (Figure~\ref{IRFTVPCPI}). We find evidence that the responsiveness of certain inflation components to monetary policy shocks has evolved over the last few decades, especially in the post-2010 period and into the post-pandemic inflation surge. Notably, the decline in durable goods inflation after a monetary tightening has become more pronounced in the recent period. In the 1990s and early 2000s, a Target shock of similar size might have produced only a modest downturn in durable goods prices, partly because inflation was low and stable and firms were reluctant to cut prices. However, by 2022–2023, after the pandemic disruptions, our model infers that a contractionary shock yields a larger immediate reduction in durable goods inflation. One interpretation is that after the pandemic’s supply-driven price spikes (especially in vehicles and goods due to supply chain issues), there was pent-up disinflationary pressure once demand slowed. So when the Fed started tightening aggressively in 2022, it popped the “demand bubble” for goods, causing a faster normalization of goods prices (indeed, used car prices, for example, started falling in late 2022). Our time-varying IRFs capture this as a bigger impact of monetary shocks on goods inflation in the recent period. 

Another dimension of heterogeneity is the peak timing of responses. Some components (energy, core goods) hit their maximum price decline within 3–6 months of the shock, whereas others (services) only show a significant change after 12–18 months. This staggered timing means that aggregate CPI inflation – a weighted sum of these – may exhibit a multi-phase reaction: an initial dip driven by energy and goods, potentially offset if energy prices rebound, and a later, more persistent slowing driven by services. This could explain why policymakers often see ``headline'' CPI inflation fall relatively quickly after tightening (as energy and food stabilize or drop), but ``core'' services inflation can remain high for a while, necessitating continued tight policy until it too comes down. Our large VAR results provide empirical backing for this narrative, which has been evident in the post-2021 episode: the Fed’s tightening in 2022 quickly lowered commodity-sensitive inflation by 2023, but core services inflation only started decelerating toward late 2023 and 2024. 

Importantly, despite these differences in timing and magnitude, virtually all CPI components eventually move in the direction of disinflation in response to a contractionary Target shock (with the possible exception of certain idiosyncratic cases like medical care prices that might be governed by institutional factors). That is, monetary policy is a blunt tool that ultimately affects broad inflation – but the path it takes through various prices is uneven. This underscores the value of a large VAR approach: it lets us observe the rich cross-sectional detail behind the aggregate responses.

\section{Conclusions}
This paper develops a novel Bayesian inference framework for identifying monetary policy shocks in large-scale VARs, addressing longstanding challenges in structural identification. By combining high-frequency surprises from financial markets with economically motivated sign restrictions, the proposed method delivers robust and interpretable structural shocks, even in high-dimensional settings. A key innovation is the integration of this hybrid identification strategy into a scalable Gibbs sampler that accommodates time-varying responses and fat-tailed disturbances, ensuring reliable inference in the presence of structural breaks and outliers.

Empirically, the paper offers new insights into the transmission of conventional monetary policy in the post-pandemic U.S. economy. By tracing the dynamic effects of interest rate surprises on disaggregated components of the consumer price index (CPI), the results highlight substantial heterogeneity in the speed and magnitude of inflation responses. While core goods prices exhibit swift and sizable declines following a rate hike, services (particularly housing) respond only gradually. These differences have become more pronounced during the 2022–24 inflationary episode, underscoring shifts in the monetary transmission mechanism and the importance of disaggregated analysis.

More broadly, the findings affirm that conventional monetary policy retains its effectiveness in shaping inflation dynamics, even in periods marked by elevated uncertainty and atypical price behavior. For policymakers, this underscores the importance of sector-specific diagnostics when assessing the stance and expected impact of policy. For researchers, the methodological framework provides a general-purpose tool for structural analysis in large systems, adaptable to a wide range of macroeconomic questions beyond monetary policy.

\newpage


\newpage

\addcontentsline{toc}{section}{References}
\bibliographystyle{apa}
\bibliography{SVAR_large}
\clearpage

\end{document}